\documentclass[12pt,journal,onecolumn]{IEEEtran}

\usepackage{amssymb,amsmath,amsfonts,bm,epsfig,graphicx,theorem,latexsym}
\usepackage{rotating,setspace,latexsym,epsf,epstopdf}
\usepackage{cite,authblk,epstopdf}

\usepackage[usenames,dvipsnames]{color}
\usepackage{soul}

\newtheorem{theorem}{Theorem}

\newtheorem{lemma}{Lemma}

\newenvironment{Proof}[1]{\medskip\par\noindent{\bf Proof:\,}\,#1}{{\mbox{\,$\blacksquare$}\medskip\par}}

\makeatletter

\IEEEoverridecommandlockouts

\def\naive{na\"{\i}ve~}

\allowdisplaybreaks[1]

\title{The Binary Energy Harvesting Channel \\ with a Unit-Sized Battery\thanks{This work was supported by NSF Grants CNS 09-64364/CNS 09-64632 and CCF 14-22347/CCF 14-22111, and presented in part at the IEEE International Symposium on Information Theory, Istanbul, Turkey, July 2013 and the IEEE International Symposium on Information Theory, Honolulu, HI, June 2014.}}

\author[1]{Kaya Tutuncuoglu}
\author[2]{Omur Ozel}
\author[1]{Aylin Yener}
\author[2]{Sennur Ulukus}
\affil[1]{\normalsize Department of Electrical Engineering, The Pennsylvania State University}
\affil[2]{\normalsize Department of Electrical and Computer Engineering, University of Maryland}

\begin{document}

\maketitle

\vspace*{-1.5cm}

\begin{abstract}
We consider a binary energy harvesting communication channel with a finite-sized battery at the transmitter. In this model, the channel input is constrained by the available energy at each channel use, which is driven by an external energy harvesting process, the size of the battery, and the previous channel inputs. We consider an abstraction where energy is harvested in binary units and stored in a battery with the capacity of a single unit, and the channel inputs are binary. Viewing the available energy in the battery as a state, this is a state-dependent channel with input-dependent states, memory in the states, and causal state information available at the transmitter only. We find an equivalent representation for this channel based on the timings of the symbols, and determine the capacity of the resulting equivalent timing channel via an auxiliary random variable. We give achievable rates based on certain selections of this auxiliary random variable which resemble lattice coding for the timing channel. We develop upper bounds for the capacity by using a genie-aided method, and also by quantifying the leakage of the state information to the receiver. We show that the proposed achievable rates are asymptotically capacity achieving for small energy harvesting rates. We extend the results to the case of ternary channel inputs. Our achievable rates give the capacity of the binary channel within 0.03 bits/channel use, the ternary channel within 0.05 bits/channel use, and outperform basic Shannon strategies that only consider instantaneous battery states, for all parameter values.
\end{abstract}

\section{Introduction} \label{sect_introduction}

We consider an energy harvesting communication channel, where the transmitter harvests energy from an exogenous source to sustain power needed for its data transmission. The transmitter stores harvested energy in a finite-sized battery, and each channel input is constrained by the remaining energy in the battery. Consequently, stored energy can be viewed as the state of this channel, which is naturally known causally at the encoder, but unknown at the decoder. This state is correlated over time, and is driven by the exogenous energy harvesting process, energy storage capacity of the battery, and the past channel inputs. As such, this channel model introduces unprecedented constraints on the channel input, departing from traditional channels with average or peak power constraints, and requires new approaches to determine its capacity.

References \cite{ozel2012achieving, ozel2011awgn, mao2013capacity, dong2014near, jog2014energy} study the capacity of channels with energy harvesting transmitters with an infinite-sized battery \cite{ozel2012achieving}, with no battery \cite{ozel2011awgn}, and with a finite-sized battery \cite{mao2013capacity, dong2014near, jog2014energy}. Reference \cite{ozel2012achieving} shows that the capacity with an infinite-sized battery is equal to the capacity with an average power constraint equal to the average recharge rate. This reference proposes save-and-transmit and best-effort-transmit schemes, both of which are capacity achieving when the battery size is unbounded. At the other extreme, \cite{ozel2011awgn} studies the case with no battery, and shows that this is equivalent to a time-varying stochastic amplitude-constrained channel. Reference \cite{ozel2011awgn} views harvested energy as a causally known state, and combines the results of Shannon on channels with causal state at the transmitter \cite{shannon1958channels} and Smith on amplitude constrained channels \cite{smith1971information}, and argues that the capacity achieving input distribution is discrete as in the case of \cite{smith1971information}. More recent work \cite{mao2013capacity, dong2014near, jog2014energy} consider the case with a finite-sized battery. Reference \cite{mao2013capacity} provides a multi-letter capacity expression that is hard to evaluate, since it requires optimizing multi-letter Shannon strategies \cite{shannon1958channels} for each channel use. The authors conjecture that instantaneous Shannon strategies are optimal for this case, i.e., strategies that only observe the current battery state to determine the channel input are sufficient to achieve the capacity. Reference \cite{dong2014near} finds approximations to the capacity of the energy harvesting channel within a constant gap of 2.58 bits/channel use. For a deterministic energy harvesting profile, \cite{jog2014energy} provides a lower bound on the capacity by exploiting the volume of energy-feasible input vectors.

\begin{figure}
\includegraphics[width=0.6\linewidth]{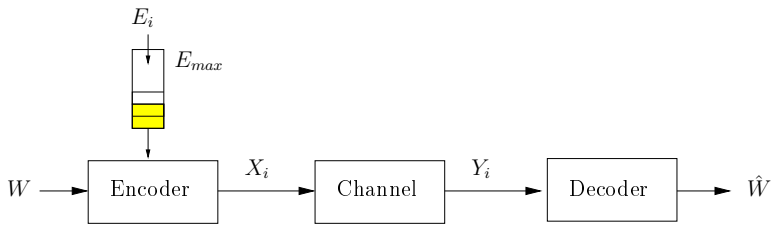}
\centering
\caption{The binary energy harvesting channel (BEHC) with an energy harvesting encoder and a finite-sized battery.}
\label{fig_model}
\end{figure}

We consider a single-user communication scenario with an energy harvesting encoder that has a finite-sized battery, as shown in Fig.~\ref{fig_model}. In each channel use, the encoder harvests energy that is a multiple of a fixed unit, and stores it in a battery which has a capacity that is also a multiple of this unit. Each channel input then consumes an integer number of units of energy. In this paper, we consider the binary version of this setting, which  we refer to as the binary energy harvesting channel (BEHC). In a BEHC, energy is harvested in binary amounts (0 or 1 unit), the battery has unit size, and the channel inputs are binary. Sending a 1 through the channel requires one unit of energy per channel use, while sending a zero is free in terms of energy. Hence, the encoder may only send a 1 when it has the required energy in the battery; it can send a 0 anytime. A similar abstraction of communicating with energy packets over an interactive link can be found in \cite{popovski2012interactive}.

In an energy harvesting channel, the channel input in each channel use is constrained by the battery state of the transmitter. Since the battery is at the transmitter, this state is naturally causally available at the encoder, but is not available at the decoder. This results in a channel with causally known state information at the encoder. In such channels, if the state is independent and identically distributed (i.i.d.) over time, and is independent of the channel inputs, then the capacity is achieved using Shannon strategies \cite{shannon1958channels}. However, in the BEHC, the battery state has memory since the battery stores the energy through channel uses. Further, the evolution of the battery state depends on the past channel inputs since different symbols consume different amounts of energy. Therefore, Shannon strategies of \cite{shannon1958channels} are not necessarily optimal for this channel. This channel model resembles the model of reference \cite{weissman2010capacity} with action dependent states, where the encoder controls the state of the channel through its own actions. However, different from \cite{weissman2010capacity}, in the case of BEHC, actions and channel inputs are equal, i.e., the two cannot be chosen independently. This yields a conflict between choosing inputs with the purpose of communicating, and with the purpose of controlling the state.

In this paper, we consider a special case of the BEHC with no channel noise. Even in this special case, finding the capacity is challenging due to the memory in the state, the lack of battery state information at the receiver, and the inter-dependence of the battery state and the channel inputs. In essence, the uncertainty in this model is not due to the communication channel, but due to the random energy harvests and the battery state that impose intricate constraints on the channel inputs. For this case, we first propose achievable rates using Shannon strategies in \cite{shannon1958channels}. Next, we develop an equivalent representation for the channel in terms of the time differences between consecutive 1s sent through the channel. This is analogous to the timing channel in \cite{anantharam1996bits}, or its discrete-time version in \cite{prabhakar2003entropy}, where the message is encoded in the arrival times of packets to a queue. Observing that the states are i.i.d.~in this equivalent representation, we find a single-letter expression for the capacity of the BEHC by combining approaches from \cite{shannon1958channels} and \cite{anantharam1996bits}. This expression is difficult to evaluate due to an involved auxiliary random variable. We give achievable rates based on certain selections of this auxiliary random variable which resemble lattice coding for the timing channel. We develop upper bounds for the capacity by using a genie-aided method, and also by quantifying the leakage of the state information to the receiver. We find that our bounds are tight asymptotically as energy harvesting rate goes to zero. We extend our results to the case of ternary channel inputs. We numerically evaluate the achievable rates and the upper bounds and show that our achievable schemes give the capacity of the binary channel within 0.03 bits/channel use and the ternary channel within 0.05 bits/channel use. We observe that the proposed timing channel based achievable schemes outperform basic Shannon strategies that consider only instantaneous battery state, for all parameter values, for this noiseless binary case.

\section{Channel Model} \label{sect_model}

We consider the binary channel with an energy harvesting transmitter shown in Fig.~\ref{fig_model}. The battery at the transmitter is of size $E_{max}$. The harvested energy is first stored in the battery before being used for transmission. The encoder transmits a symbol $X_i \in \{0,1\}$ in channel use $i$. At each channel use, the channel input $X_i$ is constrained by the energy available in the battery at that channel use. Hence, for the transmitter to send an $X_i=1$, it must have a unit of energy in the battery; the transmitter can send an $X_i=0$ anytime. Next, the encoder harvests an energy unit with probability $q$, i.e., $E_i$ is Bernoulli($q$), and stores it in its battery of size $E_{max}$ units. The harvests are i.i.d.~over time. If the battery is full, harvested energy is lost, i.e., $E_i$ cannot be used immediately in the same time slot without storing. We refer to this particular sequence of events within a channel use as the {\em transmit first} model, since the encoder first sends $X_i$ and then harvests energy $E_i$.

The battery state $S_i$ denotes the number of energy units available in the battery at the beginning of channel use $i$, and evolves as
\begin{equation}
\label{eqn_model_update}
S_{i+1}=\min \{ S_i - X_i + E_i , E_{max}\}
\end{equation}
where $X_i=0$ if $S_i=0$ due to the energy constraint. The encoder knows the battery state $S_i$ causally, i.e., at the beginning of time slot $i$, but does not know what $E_i$ or $S_{i+1}$ will be until after sending $X_i$. The decoder is unaware of the energy harvests at the encoder, and therefore the battery state. As seen from (\ref{eqn_model_update}), the battery state $S_i$ has memory, is affected by the channel inputs $X_j$ for $j\leq i$, and imposes a constraint on the channel input $X_i$. In this work, we focus on the case of a unit-sized battery, i.e., $E_{max}=1$, and a noiseless channel, i.e., $Y_i=X_i$.

\section{Achievable Rates with Shannon Strategies} \label{sect_shannon}

For a channel with i.i.d.~and causally known states at the transmitter, Shannon shows in \cite{shannon1958channels} that the capacity is achieved using the now so-called Shannon strategies. In particular, the codebook consists of i.i.d.~strategies $U_i\in \mathcal{U}$, which are functions from channel state $S_i$ to channel input $X_i$. In channel use $i$, the encoder observes $S_i$ and puts $X_i=U_i(S_i)$ into the channel. The capacity of this channel is given by
\begin{align}
\label{eqn_shannon_capacity}
	C_{CSIT}= \underset{p_U}{\max} ~ I(U;Y)
\end{align}
where $p_U$ is the distribution of $U$ over all functions from $S_i$ to $X_i$.

In the BEHC, the state of the channel, i.e., the battery state of the encoder, is not i.i.d.~over time. Therefore, (\ref{eqn_shannon_capacity}) does not give the capacity for this system. To overcome the memory in the state, \cite{mao2013capacity} uses strategies that are functions of all past battery states to express the capacity in a multi-letter form. However, since the dimension of such strategies grow exponentially with the number of channel uses, this approach is intractable. Alternatively, it is possible to use the method in \cite{shannon1958channels} to develop encoding schemes based on Shannon strategies to obtain achievable rates. One tractable such scheme is obtained when strategies are functions of the current battery state only, which is proposed as an achievable rate in \cite{mao2013capacity} and \cite{tutuncuoglu2013binary}; and is conjectured to be capacity achieving in \cite{mao2013capacity}. In this section, we consider such encoding schemes.

For the $E_{max}=1$ case, we have two states, $S_i \in \{0,1\}$. We denote a strategy $U$ as $U=(X,X^\prime)$, where $U(0)=X$ and $U(1)=X^\prime$, i.e., $X$ is the channel input when $S=0$ and $X^\prime$ is the channel input when $S=1$. Due to the inherent energy constraint of the BEHC, $X=1$ requires $S=1$, and thus, we have two feasible strategies, namely $(0,0)$ and $(0,1)$.

We first construct a codebook by choosing $U_i$ i.i.d.~for each codeword and channel use. Let the probability of choosing $U_i=(0,1)$ be $p$ for all $i$ and all codewords. We will consider two alternative approaches to decoding the message. First, note that the i.i.d.~codebook construction yields an ergodic battery state process for any message, with the transition probabilities
\begin{align}
	\mbox{Pr}[S_{i+1}=1|S_i=0]=q, \qquad \mbox{Pr}[S_{i+1}=0|S_i=1]=p(1-q)
\end{align}
yielding the stationary probability
\begin{align}
\label{eqn_shannon_naive_prob}
	\mbox{Pr}[S=1]=\frac{q}{p+q-pq}
\end{align}
The receiver can ignore the memory in the model, consider a channel with i.i.d.~states with the state probability given in (\ref{eqn_shannon_naive_prob}), and perform joint typicality decoding. This is similar to the approach used in \cite{popovski2012interactive} for a communication scenario with energy exchange. Denoting $U=(0,0)$ as $0$ and $U=(0,1)$ as $1$, this channel is expressed as
\begin{align}
\label{eqn_shannon_naive_channel}
	p(y|u)=\mbox{Pr}[S=1] \delta(y-u)+\mbox{Pr}[S=0]\delta(y)
\end{align}
where $\delta(u)$ is 1 at $u=0$, and zero elsewhere. Since the channel is memoryless, its capacity is given by  (\ref{eqn_shannon_capacity}). Note that this is an achievable rate, but it is not the capacity of the BEHC, since the decoder treats the channel as if it was memoryless. Hence, we refer to this scheme as the {\em \naive i.i.d. Shannon strategy} (NIID). The best achievable rate for the NIID scheme is given by
\begin{align}
\label{eqn_shannon_naive_rate}
	R_{NIID}=\underset{p\in[0,1]}{\max} ~ H_2 \left( \frac{pq}{p+q-pq} \right)-pH_2 \left( \frac{q}{p+q-pq} \right)
\end{align}
where $H_2(p)=-p\log(p)-(1-p)\log(1-p)$ is the binary entropy function.

While the NIID scheme permits an easy analysis, it fails to make use of the memory in the channel. Instead, the decoder can exploit the memory by using the $n$-letter joint probability $p(u^n,y^n)$ when performing joint typicality decoding. Since this is the best that can be done for an i.i.d.~codebook, we will refer to this scheme as the {\em optimal i.i.d.~Shannon strategy} (OIID), which yields the achievable rate
\begin{align}
\label{eqn_shannon_oiid_rate}
R_{OIID}=\max_{p \in [0,1]} \lim_{n \rightarrow \infty} \frac{1}{n}I(U^n;Y^n)
\end{align}
The challenge with this scheme is in calculating the limit of the $n$-letter mutual information $I(U^n;Y^n)$. To this end, we use the message passing algorithm proposed in \cite{arnold2006simulation}. This algorithm requires that the joint probability $p(y_i,u_i,s_{i+1}|s_i)$ is independent of the channel index $i$. In our case, we have independent $u_i$, which yields
\begin{align}
p(y_i,u_i,s_{i+1}|s_i)=p(y_i,s_{i+1}|u_i,s_i)p(u_i)
\end{align}
where $p(y_i,s_{i+1}|u_i,s_i)$ is independent of $i$ by the definition of the channel. Thus, we can use the algorithm in \cite{arnold2006simulation} to exhaustively search $p$ and solve (\ref{eqn_shannon_oiid_rate}).

It is possible to further improve such achievable rates by constructing more involved codebooks. For example, reference \cite{mao2013capacity} considers generating codewords with Markov processes, which introduces additional memory to the system through the codewords. This approach improves the achievable rate as shown in \cite{mao2013capacity} at the cost of increased computational complexity in the Markov order of the codebook. We evaluate and compare these achievable rates in Section~\ref{sect_numerical}.

\section{Timing Representation of the BEHC} \label{sect_timing}

In this section, we propose an alternative representation of the BEHC, which yields a simpler analysis via a single-letter expression for the capacity. In particular, we equivalently represent channel outputs $Y_i$ with the number of channel uses between instances of $Y_i=1$. We show that this transformation eliminates the memory in the state of the system, and allows constructing tractable achievable rates and upper bounds for the BEHC.

The input $X_i$ and the output $Y_i$ of the noiseless BEHC are both binary. Let $T_1 \in \{1,2,\dots\}$ be defined as the number of channel uses before the first instance of output $Y=1$, and $T_k \in \{1,2,\dots\}$ for $k \geq 2$ be defined as the number of channel uses between the $(k-1)$st instance of output $Y=1$ and the $k$th instance of output $Y=1$. In other words, the sequence $T^m$ represents the differences between the channel uses where 1s are observed at the output of the channel. Clearly, $T^m$ and $Y^n$ are equivalent since there is a unique sequence $T^m$ corresponding to each $Y^n$ and vice versa.

\begin{figure}[t]
\centerline{\includegraphics[width=0.6\linewidth]{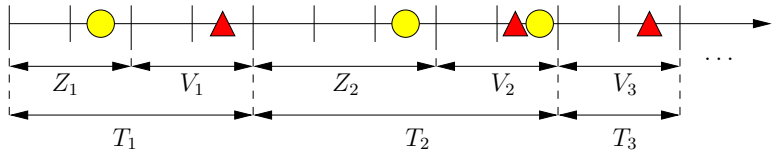}}
\caption{Graphical representation of $T_k$, $V_k$ and $Z_k$. Note that since energy is harvested immediately after sending a 1, we have $Z_3=0$.}
\label{fig_timing_model}
\end{figure}

When a 1 is transmitted in the $i$th channel use, the entire energy stored in the unit-sized battery of the encoder is consumed. Hence, the encoder cannot transmit another 1 until another energy unit is harvested. We define the idle time $Z_k \in \{0,1,\ldots\}$ of the encoder as the number of channel uses the encoder waits for energy after the $(k-1)$st 1 is transmitted. Since the probability of harvesting an energy unit is distributed i.i.d.~with Bernoulli($q$), $Z_k$ is also i.i.d.~and distributed geometric($q$) on $\{0,1,\dots\}$. Note that during the idle period, the encoder cannot send any 1s. Once the energy is harvested, the encoder observes $Z_k$ and chooses to wait $V_k \in \{1,2,\dots\}$ channel uses before sending the next 1. Hence, we have a {\em timing channel} with causally known state $Z_k$, channel input $V_k$, and channel output $T_k$, satisfying
\begin{align}
\label{eqn_timing_channel}
	T_k=V_k+Z_k
\end{align}
We illustrate the variables $T_k$, $V_k$ and $Z_k$ in Fig.~\ref{fig_timing_model}. In slots representing one use of the BEHC, an energy arrival, i.e.,  $E_i=1$, is marked with a circle and sending a 1, i.e., $X_i=1$, is marked with a triangle. Note that one use of the timing channel spans $T$ uses of the BEHC.

We remark that the timing channel constructed from the time difference between consecutive 1s resembles the noiseless channel with symbols of varying durations \cite{shannon1948}. The symbol durations in \cite{shannon1948} are fixed, while the symbol durations in our model depend on the energy harvesting process, and therefore may change each time a symbol is sent. Hence, while \cite{shannon1948} studies the problem of packing the most information within a given block length, our problem is also concerned with the randomness introduced by energy harvesting. In this sense, the timing channel defined here is analogous to the telephone signaling channel in \cite{anantharam1996bits} and its discrete time counterpart in \cite{prabhakar2003entropy}, with the exception of causal knowledge of $Z_k$ at the encoder in our model.

\subsection{Equivalence of the BEHC and the Timing Channel} \label{sub_timing_equivalence}

In the timing channel, the decoder observes $T^m$, which can be used to calculate the BEHC output sequence $Y^n$. The encoder observes $Z^m$ causally, which can be combined with past timing channel inputs $V^{m-1}$ to find the state sequence $S^n$ causally. Hence, any encoding/decoding scheme for the BEHC can be implemented in the timing channel, and vice versa, implying that the two channels are equivalent. However, note that in the timing channel, the $k$th channel use consists of $T_k$ uses of the BEHC. To take the time cost of each timing channel use into consideration, we define the timing channel capacity $C_T$ as the maximum achievable message rate per use of the BEHC channel. In particular, given a timing channel codebook consisting of $M$ codewords of length $m$, sending a codeword takes $n=m\mathbb{E}[T]$ uses of the BEHC on average, and the corresponding rate is defined as
\begin{align}
\label{eqn_timing_scaled_rate}
R=\frac{\log M}{m \mathbb{E}[T]}=\frac{\log M}{n}
\end{align}
We remark that this definition is a variation of the rate of the telephone signaling channel introduced in \cite[Defn.~5]{anantharam1996bits}. With both rates defined per use of the binary channel, the timing channel and the BEHC have the same capacity. This is due to the encoders and decoders of these channels having different but equivalent representations of the same channel. We state this fact as a lemma.

\begin{lemma}
\label{lem_timing_equivalent}
The timing channel capacity with additive causally known state at the encoder, $C_{T}$, and the BEHC capacity, $C_{BEHC}$, are equal, i.e., $C_{BEHC}=C_T$.
\end{lemma}

\subsection{Capacity of the Timing Channel} \label{sub_timing_capacity}

The timing channel defined in (\ref{eqn_timing_channel}) is memoryless since $Z_k$ are independent. For such channels, the capacity is given by (\ref{eqn_shannon_capacity}), or more explicitly by the following expression \cite{shannon1958channels}
\begin{equation}
\label{eqn_timing_csit}
	C_{CSIT}=\underset{p(u),v(u,z)}{\max}~ I(U;T)
\end{equation}
where $U$ is an auxiliary random variable that represents the Shannon strategies, and $v(U,Z)$ is a mapping from auxiliary $U$ and state $Z$ to the channel input $V$. The cardinality bound on the auxiliary random variable is $|\mathcal{U}| \leq \min \{(|\mathcal{V}|-1)|\mathcal{Z}|+1,|\mathcal{T}|\}$. As stated in \cite[Thm.~7.2]{el2011network}, a deterministic $v(u,z)$ can be assumed without losing optimality. Hence, solving (\ref{eqn_timing_csit}) requires finding the optimal distribution for $U$, $p(u)$, and the optimal deterministic mapping $v(u,z)$.

Due to Lemma~\ref{lem_timing_equivalent}, we are interested in $C_T$, which is defined per use of the binary channel, i.e., with a time cost of $T_k$ for the $k$th channel use. To this end, we combine the approaches in \cite{shannon1958channels} for channels with causal state information at the transmitter, and \cite{anantharam1996bits} for timing channels, to state the following theorem.

\begin{theorem}
\label{thm_timing_capacity}
The capacity of the timing channel with additive causally known state, $C_{T}$, is
\begin{equation}
\label{eqn_timing_capacity}
	C_{T}=\underset{p(u),v(u,z)}{\max}~ \frac{I(U;T)}{\mathbb{E}[T]}
\end{equation}
\end{theorem}

\begin{Proof}
Let $W$ denote the message which is uniform on $\{1,\ldots,M\}$. Let $n$ be the maximum number of binary channel uses, averaged over the energy arrivals $E_i$, to send a message $W=w$. We note that by definition, we have
\begin{align}
\label{eqn_timing_proof1}
\sum_{k=1}^{m} \mathbb{E}[T_k] \leq n
\end{align}
where the expectation is over the energy arrival sequence $E_i$ and the message $W$.

For the converse proof, we define $U_k = (W,T^{k-1})$. Since $E_i$ is an i.i.d.~random process, $Z_k$ is independent of $W$ and $T^{k-1}$, and therefore $U_k$. We write
\begin{align}
\label{eqn_timing_conv_0}
\log(M) - H(W|T^m) &= H(W) - H(W|T^m) \\
\label{eqn_timing_conv_1}
&= I(W;T^{m}) \\
\label{eqn_timing_conv_2}
&= \sum_{k=1}^{m} I(W;T_k | T^{k-1}) \\
\label{eqn_timing_conv_3}
&\leq \sum_{k=1}^m I(W,T^{k-1};T_k) \\
\label{eqn_timing_conv_4}
&= \sum_{k=1}^m I(U_k;T_k) \\
\label{eqn_timing_conv_5}
&\leq \frac{n}{\sum_{k=1}^m \mathbb{E}[T_k]} \sum_{k=1}^m I(U_k;T_k) \\
\label{eqn_timing_conv_6}
&\leq n \sup_{U} \frac{I(U;T)}{\mathbb{E}[T]} =n C_T
\end{align}
where (\ref{eqn_timing_conv_5}) follows from (\ref{eqn_timing_proof1}), and (\ref{eqn_timing_conv_6}) follows from $U_i$ being independent of $Z_i$ and the inequality $\frac{\sum_i a_i}{\sum_i b_i} \leq \max_i \frac{a_i}{b_i}$, for $a_i,b_i >0$. When $m \rightarrow \infty$, if the probability of error goes to zero, then Fano's inequality implies $H(W|T^m) \rightarrow 0$. Combining this with (\ref{eqn_timing_scaled_rate}) and (\ref{eqn_timing_conv_6}), we get $\frac{\log(M)}{n}=R \leq C_T$, which completes the converse proof.

For the achievability of this rate, we use the encoding scheme in \cite{shannon1958channels}. In particular, the message rate $I(U;T)$ per use of the timing channel is achievable with a randomly generated codebook consisting of strategies $U_k$ \cite{shannon1958channels}. Therefore, as $m \rightarrow \infty$, we have $n=m\mathbb{E}[T]$, and the message rate $R=\frac{I(U;T)}{\mathbb{E}[T]}$ per use of the BEHC is achievable, completing the achievability proof.
\end{Proof}

We noted in Section~\ref{sect_shannon} that the optimal distribution over Shannon strategies can be found numerically for the BEHC. This is due to the fact that for a binary input $X_i$ and binary state $S_i$, there are only two feasible Shannon strategies. However, for the timing channel, both the input $V_k \in \{1,2,\dots\}$ and the state $Z_k \in \{0,1,\dots\}$ have infinite cardinalities. This also implies that the cardinality bound on $U$ is infinite. Therefore, although (\ref{eqn_timing_capacity}) is a single-letter expression, it is difficult to evaluate explicitly. In the following sections, we first develop upper bounds for the capacity using a genie-aided method and using a method that quantifies the leakage of the state information to the receiver; and then develop lower bounds (explicit achievable schemes) by certain specific selections for $p(u)$ and $v(u,z)$; and compare these achievable rates and the upper bounds.

\section{Upper Bounds on the Capacity of the BEHC} \label{sect_upperbounds}

\subsection{Genie Upper Bound} \label{sub_ub_genie}

We first provide the timing channel state $Z_k$ to the decoder as genie information. This yields an upper bound since the decoder can choose to ignore $Z_k$ in decoding. However, with the knowledge of $Z_k$, the decoder can calculate $V_k=T_k-Z_k$, and thus we obtain the upper bound
\begin{align}
\label{eqn_ub_genie_1}
	C_{UB}^{genie} &= \underset{p(v)}{\max}~ \frac{H(V)}{\mathbb{E}[V]+\mathbb{E}[Z]} \\
\label{eqn_ub_genie_2}
			       &=\underset{\mu \geq 1}{\max}~ \frac{1}{\mu+\mathbb{E}[Z]} ~ \underset{\mathbb{E}[V] \leq \mu}{\max}~ H(V)
\end{align}
Note that in (\ref{eqn_ub_genie_2}), we partition the maximization into choosing the optimal $\mathbb{E}[V]=\mu$ and choosing the optimal distribution of $V$ with $\mathbb{E}[V] \leq \mu$. The equality in (\ref{eqn_ub_genie_2}) holds since the term $(\mu+\mathbb{E}[Z])^{-1}$ is decreasing in $\mu$, and therefore the optimal $\mu$ equals the expectation of the optimal $V$. The second maximization in (\ref{eqn_ub_genie_2}) involves finding the entropy maximizing probability distribution over the discrete support set $\mathbb{Z}^+=\{1,2,\dots\}$ with the constraint $\mathbb{E}[V] \leq \mu$. The solution to this problem is a geometric distributed $V$ with parameter $\frac{1}{\mu}$. Its entropy is given by $H(V)=\frac{H_2(p)}{p}$, where $H_2(p)$ is the binary entropy function. Noting that $Z$ is also geometrically distributed with parameter $q$, the genie upper bound reduces to
\begin{equation}
\label{eqn_ub_genie_final}
	C_{UB}^{genie} =\underset{p \in [0,1]}{\max}~ \frac{H_2(p)/p}{\frac{1}{p}+\frac{1-q}{q}} = \underset{p \in [0,1]}{\max}~ \frac{qH_2(p)}{q+p(1-q)}
\end{equation}

The genie upper bound in (\ref{eqn_ub_genie_final}) overcomes the state dependence of the timing channel by effectively removing the state $Z_k$ from the channel. Although this neglects the main challenges of our model, we will show in Section~\ref{sub_achievable_asymptotic} that this is a useful upper bound which in fact is asymptotically optimal as $q \rightarrow 0$.

\subsection{State Leakage Upper Bound} \label{sub_ub_leakage}

Another approach to obtain an upper bound is to quantify the minimum amount of information $T^m$ carries about $Z^m$. Since $Z^m$ is independent of the message, information {\em leaked} about it via $T^m$ reduces the potential information that can be carried in $T^m$ about the message. Following this intuition, in this subsection, we find an upper bound on $H(Z|T=t,U=u)$, which yields the state leakage upper bound for the timing channel capacity.

An example that relates to this idea can be found in \cite{tavan2013bits}. This reference considers communicating through a queue with a single packet buffer, where the encoding is performed over arrival times to the buffer. The decoder recovers the message by observing the buffer departure times of packets, which have suffered random delays through the buffer. What this example suggests is that it is possible to achieve a positive message rate through a buffer that causes random delays. In a similar manner, we can consider timing channel input $V$ as random delay, and achieve a positive rate between the harvesting process and the decoder in addition to the message rate of the timing channel. Since the total message rate is limited to $H(Y)$ or $H(T)/\mathbb{E}[T]$ by the cutset bound, quantifying this nonzero rate between the harvesting process and the decoder is useful in finding an upper bound.

We first present the following lemma, where we provide an upper bound for $H(Z|T=t,U=u)$. This conditional entropy represents the amount of uncertainty remaining in $Z$ after the decoder receives $T$ and successfully decodes $U$.

\begin{lemma}
\label{lem_ub_truncated}
For the timing channel $T=V+Z$, where $Z$ is geometric with parameter $q$, and $V=v(U,Z)$ with the auxiliary random variable $U$ independent of $Z$, we have
\begin{equation}
\label{eqn_ub_leakage_lemma}
H(Z|T=t,U=u) \leq H(Z_t)
\end{equation}
where $Z_t$ is a truncated geometric random variable on $\{0,1,\ldots,t-1\}$ with the probability mass function
\begin{equation}
\label{eqn_ub_leakage_pzt}
p_{Z_t}(z)=
\begin{cases}
\frac{q(1-q)^z}{1-(1-q)^t}, & \mbox{if}~z<t \\
0,			& \mbox{otherwise}
\end{cases}
\end{equation}
\end{lemma}
\begin{Proof}
We first examine the joint distribution $p(z,t|u)$ resulting from a deterministic $v(U,Z)$, which is depicted as a two-dimensional matrix in Fig.~\ref{fig_ub_leakage_lemma}. Given $Z=z$ and $U=u$, the output of the channel is $T=v(u,z)+z$. Therefore, each row of $p(z,t|u)$ in the figure contains one non-zero term. We also have
\begin{align}
\label{eqn_ub_leakage_l0}
p(z,t|u)=0, \qquad z \geq t
\end{align}
since $v(u,z)$ is positive by definition. This is denoted by the shaded area in the figure. Moreover, we write
\begin{align}
\label{eqn_ub_leakage_l1}
p(z,v(u,z)+z|u)	&= \sum_{t=1}^\infty p(z,t|u)\\
\label{eqn_ub_leakage_l2}
				&= p(z|u) = p(z)
\end{align}
implying that the non-zero term in row $z$ is equal to $\mbox{Pr}[Z=z]$. Here, the second equality in (\ref{eqn_ub_leakage_l2}) follows from the independence of $U$ and $Z$.

\begin{figure}
\includegraphics[width=0.6\linewidth]{figures/xfig/lemma.eps}
\centering
\caption{The joint probability matrix $p(z,t|u)$ for a fixed strategy $u$. There is one non-zero term in each row, which equals $p(z)$. When calculating $H(Z|T=t,U=u)$, only the values in the bold rectangle are required.}
\label{fig_ub_leakage_lemma}
\end{figure}

To find $H(Z|T=t,U=u)$, we focus on column $t$ of the probability matrix $p(z,t|u)$, which is marked with a bold rectangle in the figure. Let $\mathcal{A} \subset \{0,1,\ldots,t-1\}$ denote the set of indices $z \in \{0,1,\ldots\,t-1\}$ for which $p(z,t|u)=p(z)$. As such, we can write $p(z|t,u)$ as
\begin{align}
\label{eqn_ub_leakage_l3}
	p_{\mathcal{A}}(z)=p(z|t,u)	&= \frac{p(z,t|u)}{\sum_{t=1}^\infty p(z,t|u)} \\
			&=
			\begin{cases}
				\frac{q(1-q)^z}{\sum_{a\in \mathcal{A}} q(1-q)^{a}}, &\mbox{if}~ z\in \mathcal{A} \\
				0,	& \mbox{otherwise}
			\end{cases}
\end{align}

We next prove that $H(Z|T=t,U=u)$ is maximized when $\mathcal{A}^*=\{0,1,\ldots,t-1\}$, i.e., when all terms in the bold rectangle in Fig.~\ref{fig_ub_leakage_lemma} are non-zero. To this end, we show that the distribution $p_{\mathcal{A}^*}(z)$ is majorized by $p_{\mathcal{A}}(z)$ for all index sets  $\mathcal{A}=\{a_0,a_1,\ldots,a_{k-1}\} \subset \{0,1,\ldots,t-1\}$, $k \leq t$. Without loss of generality, we assume that $a_0<a_1<\ldots<a_{k-1}$, which implies the ordering
\begin{align}
	p_{\mathcal{A}}(a_0) > p_{\mathcal{A}}(a_1) > ... > p_{\mathcal{A}}(a_{k-1})
\end{align}
for any $\mathcal{A}$. For $0 \leq n \leq k-1$, we write
\begin{align}
\label{eqn_ub_leakage_lp1}
\sum_{i=0}^n p_{\mathcal{A}}(a_i) 	&= \frac{\sum_{i=0}^n q(1-q)^{a_i}}{\sum_{i=0}^{k-1} q(1-q)^{a_i}} \\
\label{eqn_ub_leakage_lp3}
&\geq \frac{\sum_{i=0}^n (1-q)^{a_n+i-n}}{\sum_{i=0}^{n} (1-q)^{a_n+i-n} + \sum_{i=n+1}^{k-1} (1-q)^{a_i}} \\
\label{eqn_ub_leakage_lp5}
&\geq \frac{\sum_{i=0}^n (1-q)^{a_n+i-n}}{\sum_{i=0}^{k-1} (1-q)^{a_n+i-n}}\\
\label{eqn_ub_leakage_lp7}
& \geq \frac{\sum_{i=0}^n (1-q)^i}{\sum_{i=0}^{t-1} (1-q)^i} = \sum_{i=0}^n p_{\mathcal{A}^*}(i)
\end{align}
where we obtain (\ref{eqn_ub_leakage_lp3}) by subtracting
\begin{equation}
\delta_1 = \sum_{i=0}^n (1-q)^{a_i} - \sum_{i=0}^n (1-q)^{a_n+i-n}
\end{equation}
from both the numerator and the denominator, and we obtain (\ref{eqn_ub_leakage_lp5}) by adding
\begin{equation}
\delta_2 = \sum_{i=n+1}^{k-1} (1-q)^{a_n+i-n} - \sum_{i=n+1}^{k-1} (1-q)^{a_i}
\end{equation}
to the denominator. Note that both $\delta_1$ and $\delta_2$ are non-negative since $a_n-a_i \geq n-i$, for $n \geq i$. Finally, (\ref{eqn_ub_leakage_lp7}) follows from $k \leq t$.

Due to the concavity of $f(x)=-x \log(x)$, and since the set $\mathcal{A}$ is finite, the majorization shown in (\ref{eqn_ub_leakage_lp1})-(\ref{eqn_ub_leakage_lp7}) implies that $H(Z|T=t,U=u)$ is maximized for $\mathcal{A}^*=\{0,1,\ldots,t-1\}$. In this case, the conditional distribution of $Z$ given $t$ and $u$ is truncated geometric. Hence, for any $v(U,Z)$, $H(Z|T=t,U=u)$ is upper bounded by the entropy of a truncated geometric random variable, $H(Z_t)$.
\end{Proof}

Using the bound obtained in Lemma~\ref{lem_ub_truncated}, we next present the leakage upper bound on the timing channel capacity $C_T$.

\begin{theorem}
\label{thm_ub_leakage}
The capacity of the timing channel and therefore the BEHC is upper bounded by
\begin{align}
\label{eqn_ub_leakage_upperbound}
C_{UB}^{leakage} = \underset{p_T(t) \in \mathcal{P}}{\max}~ \frac{H(T)-\sum_{t=1}^\infty \frac{H_2((1-q)^t)}{1-(1-q)^t}p(t)}{\mathbb{E}[T]}
\end{align}	
where $H_2(\cdot)$ is the binary entropy function, and
\begin{align}
\label{eqn_ub_leakage_p}
\mathcal{P} = \left\{ p_T(t)  \bigg| \sum_{t=1}^s p(t) \leq 1-(1-q)^s, ~ s=1,2,\ldots  \right\}
\end{align}
\end{theorem}

\begin{Proof}
Using the chain rule of mutual information, we write the numerator of (\ref{eqn_timing_capacity}) as
\begin{align}
I(U;T)	&= I(U,Z;T)-I(Z;T|U) \\
&= H(T)-H(T|U,Z) - I(Z;T|U) \\
\label{eqn_ub_leakage_thm1} &= H(T) - I(Z;T|U)
\end{align}
where the last equality follows since $T=v(U,Z)+Z$ is a deterministic function of $U$ and $Z$. Note that the $I(Z;T|U)$ term in (\ref{eqn_ub_leakage_thm1}) quantifies the information leaked to the decoder about the energy harvesting process $Z$. We lower bound this term as
\begin{align}
I(Z;T|U) &= H(Z|U) - H(Z|T,U) \\
\label{eqn_ub_leakage_thm2}
& = H(Z) - H(Z|T,U) \\
& = \sum_{t=1}^\infty \sum_{u} p(t,u)\left[ H(Z) - H(Z|T=t,U=u) \right] \\
\label{eqn_ub_leakage_thm3}
& \geq \sum_{t=1}^\infty \left[ H(Z) - H(Z_t) \right] \sum_{u} p(t,u) \\
\label{eqn_ub_leakage_thm4}
& = \sum_{t=1}^\infty \left[ H(Z) - H(Z_t) \right] p(t)
\end{align}
where (\ref{eqn_ub_leakage_thm2}) is due to the independence of $Z$ and $U$, and (\ref{eqn_ub_leakage_thm3}) is due to Lemma~\ref{lem_ub_truncated}. Substituting (\ref{eqn_ub_leakage_thm1}) and (\ref{eqn_ub_leakage_thm4}) in (\ref{eqn_timing_capacity}), we get
\begin{align}
\label{eqn_ub_leakage_thm5}
C_T \leq \underset{p(u),v(u,z)}{\max} \frac{H(T)-\sum_{t=1}^\infty [H(Z)-H(Z_t)] p(t)}{\mathbb{E}[T]}
\end{align}
Note that the objective is a function of $p_T(t)$ only. Therefore, without loss of generality, we can perform the maximization over distributions $p_T(t)$ that are achievable by some auxiliary $p_U(u)$ and function $v(U,Z)$. Since $T>Z$ by definition, such a distribution must satisfy
\begin{align}
\label{eqn_ub_leakage_thm6}
\sum_{t=1}^s p(t) \leq \sum_{z=0}^{s-1} p(z) = 1-(1-q)^s, \quad s=1, 2, \ldots
\end{align}
As a result, the distribution $p_T(t)$ induced by any $p_U(u)$ and $v(U,Z)$ lies in the set of distributions $\mathcal{P}$ defined in (\ref{eqn_ub_leakage_p}). We finally note that for geometrically distributed $Z$ and truncated geometric distributed $Z_t$, we have
\begin{align}
\label{eqn_ub_leakage_thm7}
H(Z) - H(Z_t) = \frac{H_2((1-q)^t)}{1-(1-q)^t}
\end{align}
Substituting (\ref{eqn_ub_leakage_thm6}) and (\ref{eqn_ub_leakage_thm7}) in (\ref{eqn_ub_leakage_thm5}), we arrive at the upper bound in (\ref{eqn_ub_leakage_upperbound})-(\ref{eqn_ub_leakage_p}).
\end{Proof}

\subsection{Computing the State Leakage Upper Bound} \label{sub_ub_computing}

Solving (\ref{eqn_ub_leakage_upperbound}) requires finding the optimal $p(t)$ distribution in $\mathcal{P}$. We next find the properties of the optimal distribution $p^*(t)$ to simplify its calculation. We begin by rewriting the maximization problem in (\ref{eqn_ub_leakage_upperbound}) as
\begin{align}
\label{eqn_ub_calculating_1}
	C_{UB}^{leakage} = \underset{\beta}{\max}~~ \frac{1}{\beta} ~~ \underset{p_T(t) \in \mathcal{P}, \mathbb{E}[T] \leq \beta }{\max} H(T)-\sum_{t=1}^\infty \Delta_t p(t)
\end{align}	
where we have defined $\Delta_t=\frac{H_2((1-q)^t)}{1-(1-q)^t}$. The inner maximization in (\ref{eqn_ub_calculating_1}) is a convex program since it has a concave objective and linear constraints. For this problem, we write the KKT optimality conditions\cite{bertsekas1999nonlinear} as
\begin{align}
	\label{eqn_ub_calculating_stat1}
p(t) = \exp \left(-\mu t-\Delta_t + \lambda_t - \sum_{s=1}^t \gamma_s - \eta - 1 \right) &, \quad t=1,2,\ldots \\
	\label{eqn_ub_calculating_cs1}
\lambda_t p(t) =0&, \quad \lambda_t \geq 0 \\
	\label{eqn_ub_calculating_cs2}
\gamma_t \left( \sum_{s=1}^t p_T(s) - 1 + (1-q)^t \right) =0&, \quad \gamma_t \geq 0 \\
	\label{eqn_ub_calculating_cs3}
\mu \left( \mathbb{E}[T] - \beta \right) =0&, \quad \mu \geq 0 \\
	\label{eqn_ub_calculating_cs4}
\eta \left( \sum_{s=1}^\infty p_T(s) - 1 \right) = 0&
\end{align}
where $\lambda_t$, $\gamma_t$, $\mu$ and $\eta$ are the Lagrange multipliers for the constraints $p(t) \geq 0$, $\sum_{s=1}^t p_T(s) \leq 1-(1-q)^t$, $\mathbb{E}[T] \leq \beta$, and $\sum_{s=1}^{\infty} p_T(s) = 1$, respectively.

In order to have $p(t)=0$ for some $t$, we need the exponent term in (\ref{eqn_ub_calculating_stat1}) to go to $-\infty$. This makes $\lambda_t$ in the expression of $p(t)$ redundant due to (\ref{eqn_ub_calculating_cs1}). Hence, we assign $\lambda_t=0$ for all $t$, and obtain
\begin{align}
\label{eqn_ub_calculating_stat2}
	p^*(t) = A \exp \left( -\mu t-\Delta_t-\sum_{n=1}^t \gamma_n \right)
\end{align}	
where we have defined $A= e^{- \eta - 1}$. We find $A$ from (\ref{eqn_ub_calculating_cs4}) for all $\mu \geq 0$ and $\gamma_i$ as
\begin{align}
\label{eqn_ub_calculating_stat3}
	A = \left( \sum_{t=1}^\infty e^{-\mu t-\Delta_t-\sum_{n=1}^t \gamma_n} \right)^{-1}
\end{align}
which, together with (\ref{eqn_ub_calculating_stat2}), gives us a class of distributions with parameters $\gamma_t$ and $\mu$. In addition, from (\ref{eqn_ub_calculating_cs2}), we know that $\gamma_t$ is positive only when the constraint in (\ref{eqn_ub_leakage_thm6}) is satisfied with equality. As a result, for each value of $\beta$, we can find the optimal distribution $p^*(t)$ numerically by searching the class of distributions in (\ref{eqn_ub_calculating_stat2}) for the optimal $\gamma_t$ and $\mu$ satisfying the above conditions.

\section{Achievable Rates for the BEHC} \label{sect_achievable}

In this section, we propose two choices for the auxiliary random variable $U$ and the mapping $v(u,z)$ in (\ref{eqn_timing_capacity}) and find lower bounds on the timing channel capacity and hence the BEHC capacity.

\subsection{Modulo Encoding with Finite Cardinality Auxiliary Random Variables} \label{sub_achievable_modulo}

Let $U$ be distributed over the finite support set $\{0,1,\ldots,N-1\}$, where $N$ is a parameter to be optimized. We choose the mapping
\begin{equation}
\label{eqn_achievable_modulo_v}
	v(U,Z)=(U-Z \mbox{ mod } N) + 1
\end{equation}
which gives a channel input $V=v(U,Z)$ in $\{1,2,\dots,N\}$. The output of the timing channel becomes $T=V+Z=(U-Z \mbox{ mod } N) + 1 + Z$. The decoder calculates
\begin{align}
	T^\prime &= (T-1 \mbox{ mod } N) = ((U-Z \mbox{ mod } N) + Z \mbox{ mod } N) \\
	&= U \mbox{ mod } N = U
\end{align}
and therefore perfectly recovers $U$ in each channel use. Hence, the achievable rate for this $N$ is
\begin{align}
\label{eqn_achievable_modulo_rn}
R_{A}^{(N)} = \underset{p(u),~U\in\{0,\dots,N-1\}}{\max}~ \frac{H(U)}{\mathbb{E}[V+Z]}
\end{align}
We then find the best rate achievable with this scheme by optimizing over $N$ as
\begin{align}
\label{eqn_achievable_modulo_r}
R_{A}^{mod}=\underset{N}{\max}~ R_{A}^{(N)}
\end{align}

\begin{figure}[t]
\centerline{\includegraphics[width=0.6\linewidth]{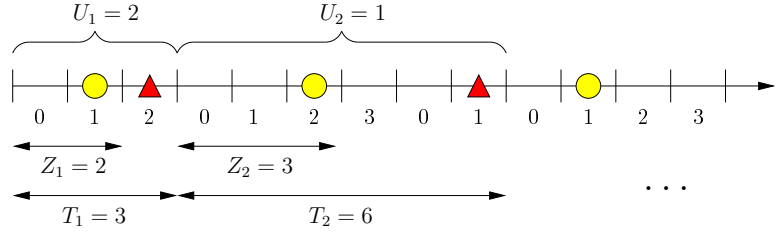}}
\caption{Modulo encoding: each message symbol $U_i$ is conveyed by transmitting a $1$ at the earliest channel use possible with index equal to $U_i$. Here, $N=4$.}
\label{fig_achievable_modulo}
\end{figure}

This encoding scheme has the following interpretation for the BEHC: Consider that after each instance of $X_i=1$, future channel uses are indexed cyclically with the numbers $\{0,1,\ldots,N-1\}$, as illustrated in Fig.~\ref{fig_achievable_modulo} for $N=4$. These indices are available to both the encoder and the decoder since the channel is noiseless. The encoder can then convey any symbol $U\in\{0,1,\ldots,N-1\}$ to the decoder by sending a 1 in a channel use indexed with $U$. This is performed at the earliest possible such channel use in which the required energy is available. For example, $U_1=2$ in the figure is conveyed in the first channel use indexed with a 2 (in the first frame of $N$ channel uses) as the energy becomes available for that transmission. However, $U_2=1$ in the figure is conveyed in the second channel use indexed with a 1 (in the second frame of $N$ channel uses), since energy is not yet harvested in the first channel use indexed by a 1 (in the first frame of $N$ channel uses). As such, in this coding scheme, the encoder partitions future channel uses into frames of length $N$, and uses the earliest feasible frame to convey its symbol $U_k$.

This encoding scheme resembles the idea of \emph{concentration} proposed by Willems in \cite{willems1988gaussian, willems2000signaling} for Gaussian channels with causal state information. In particular, part of the channel input in \cite{willems1988gaussian, willems2000signaling} is used to concentrate the channel state onto a set of values so that it can be decoded and eliminated at the decoder. Here, by waiting for the next frame of length $N$ when necessary, the effective state $Z_k$ is concentrated onto the lattice of the integer multiples of $N$. The concentrated state is then removed by the decoder with the modulo operation when calculating $T^\prime$. Hence, this encoding scheme can also be interpreted as \emph{lattice-coding} in the timing channel.

\subsection{Asymptotic Optimality of Modulo Encoding} \label{sub_achievable_asymptotic}

We next show that the modulo encoding scheme proposed in Section~\ref{sub_achievable_modulo} is asymptotically optimal as the harvest rate $q \rightarrow 0$. We establish this by comparing the achievable rate of the modulo encoding scheme in (\ref{eqn_achievable_modulo_rn})-(\ref{eqn_achievable_modulo_r}) with the genie-aided upper bound in (\ref{eqn_ub_genie_final}).

\begin{theorem}
\label{thm_asymptotic}
The modulo encoding scheme for the timing channel with auxiliary $U\in\{0,1,\dots,N-1\}$ and the channel input given in (\ref{eqn_achievable_modulo_v}) is asymptotically optimal as energy harvest rate $q \rightarrow 0$.
\end{theorem}
\begin{Proof}
We show that the upper bound $C_{UB}^{genie}$ and the achievable rate $R_A^{mod}$ scale with the same rate as $q$ goes to zero, i.e.,
\begin{align}
\label{eqn_achievable_asymptotic_thm1}
\underset{q \rightarrow 0}{\lim} ~\frac{C_{UB}}{R_A} = 1
\end{align}
For fixed $q$, the problem in (\ref{eqn_ub_genie_final}) is convex since the objective is continuous, differentiable, and concave in $p$. Therefore, the optimal $p^*$ solving (\ref{eqn_ub_genie_final}) is the solution of
\begin{align}
\frac{q(\log(1-p^*)-q\log(p^*))}{(p^*+q-p^*q)^2} = 0
\end{align}
which reduces to
\begin{align}
\label{eqn_achievable_asymptotic_thm2}
q=\frac{\log(1-p^*)}{\log(p^*)}
\end{align}
for $q>0$. Consequently, there exists an optimal $0 < p^* \leq 0.5$ for all harvest rates $0 < q \leq 1$, which approaches zero with $q$, i.e.,
\begin{equation}
\label{eqn_achievable_asymptotic_thm3}
\underset{q \rightarrow 0}{\lim} ~p^* = 0
\end{equation}

We choose the parameters of the encoding scheme as $N=\left \lceil \frac{1}{p^*} \right \rceil$, and $p(u)=1/N$ for $0 \leq u \leq N-1$, i.e., $U$ is uniformly distributed. Note that $p^* \leq 0.5$ implies $N\geq 2$. Since $U$ is uniform and independent of $Z$, from (\ref{eqn_achievable_modulo_v}), we observe that $V$ is distributed uniformly on $\{1,2,\dots,N\}$. This gives $\mathbb{E}[V]=(N+1)/2$, and the achievable rate for this scheme becomes
\begin{align}
\label{eqn_achievable_asymptotic_thm4}
R_{A}^{mod} &= \frac{H(U)}{\mathbb{E}[V]+\mathbb{E}[Z]} = \frac{\log(N)}{\frac{N+1}{2}+\frac{1-q}{q}} \geq \frac{q\log(N)}{Nq+1-q}
\end{align}
where $\mathbb{E}[Z]=(1-q)/q$. Observing that the last term in (\ref{eqn_achievable_asymptotic_thm4}) is increasing in $N$ within the interval $[\frac{1}{p^*},\lceil \frac{1}{p^*} \rceil]$, we further lower bound $R_{A}^{mod}$ as
\begin{align}
\label{eqn_achievable_asymptotic_thm5}
R_{A}^{mod} \geq \frac{q\log(N)}{Nq+1-q} \geq \frac{-qp^*\log(p^*)}{q+p^*(1-q)} = \bar{R}_A
\end{align}
and upper bound the left hand side of (\ref{eqn_achievable_asymptotic_thm1}) as
\begin{align}
\label{eqn_achievable_asymptotic_thm6}
\underset{q \rightarrow 0}{\lim} ~\frac{C_{UB}^{genie}}{R_A^{mod}} &\leq \underset{q \rightarrow 0}{\lim} ~\frac{C_{UB}^{genie}}{\bar{R}_A} \\
\label{eqn_achievable_asymptotic_thm7}
				&=\underset{q \rightarrow 0}{\lim} ~ \frac{qH(p^*)}{q+p^*(1-q)} \cdot \frac{q+p^*(1-q)}{-qp^*\log(p^*)} \\
\label{eqn_achievable_asymptotic_thm8}
				&= 1 + \underset{p^* \rightarrow 0}{\lim} ~ \frac{(1-p^*)\log(1-p^*)}{p^*\log(p^*)} = 1
\end{align}
Since $C_{UB}^{genie} \geq R_A^{mod}$ by definition, this proves (\ref{eqn_achievable_asymptotic_thm1}) and thus the theorem.
\end{Proof}

Theorem~\ref{thm_asymptotic} states that as $q \rightarrow 0$, the capacity achieving encoding scheme approaches a uniformly distributed $U$ over $\{0,\ldots,N-1\}$, where $N \rightarrow \infty$. This gives us a simple and asymptotically optimal encoding scheme for scenarios with very low energy harvesting rates.

\subsection{Extended Modulo Encoding} \label{sub_achievable_extended}

To improve the rates achievable with modulo encoding of Section~\ref{sub_achievable_modulo}, we propose an extended version of the scheme with $U\in\{0,1,\ldots\}$ and
\begin{align}
\label{eqn_achievable_extended_v}
v(U,Z)=
\begin{cases}
U-Z + 1, 					& U \geq Z \\
(U-Z \mbox{ mod } N) + 1,  	& U < Z
\end{cases}
\end{align}

\begin{figure}[t]
\centerline{\includegraphics[width=0.6\linewidth]{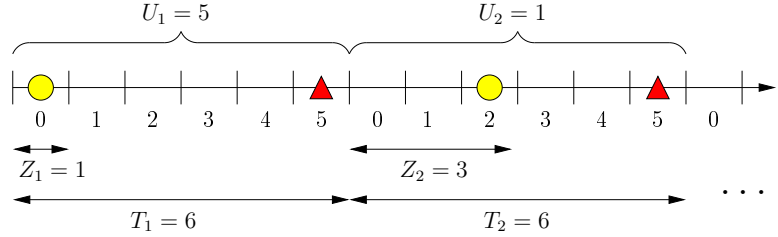}}
\caption{Extended modulo encoding for $N=4$.}
\label{fig_achievable_extended}
\end{figure}

The interpretation of this encoding scheme for the BEHC is given in Fig.~\ref{fig_achievable_extended} for $N=4$. Unlike modulo encoding, we index channel uses with $\{0,1,\dots\}$ in this case. If the required energy is harvested by the channel use indexed with $U_k$, then the encoder sends a 1 in that channel use, as is the case for $U_1$ in the figure. However, if the intended channel use is missed due to lack of energy, the encoder sends a 1 within $N$ channel uses after harvesting energy, such that the channel index and $U_k$ are equal in modulo $N$. An example is $U_2$ in the figure, where the channel index and $U_2$ are equal in modulo $N$, i.e.,
\begin{align}
(T - 1) \mbox{ mod }N= U \mbox{ mod }N,
\end{align}

The achievable rate for this scheme is calculated by solving
\begin{align}
\label{eqn_achievable_extended_r}
	R_{A}^{ext} &=\underset{N}{\max}~ \underset{p(u),~U\in\{0,1,\dots\}}{\max}~ \frac{I(U;Y)}{\mathbb{E}[V+Z]}
\end{align}
numerically by searching distributions of $U$. Although this problem is more difficult than that in (\ref{eqn_achievable_modulo_r}), it is more tractable than (\ref{eqn_timing_capacity}) since the function $v(U,Z)$ is fixed.

We note that this scheme is an extended version of the modulo encoding scheme in Section~\ref{sub_achievable_modulo}, where $U$ is not restricted to be within $[0,N-1]$. Therefore, the extended modulo scheme also includes the modulo scheme as a special case when $p(u)=0$ for $u \geq N$. In fact, this scheme can be interpreted as a combination of modulo encoding and a {\em best effort} encoding scheme where the closest feasible symbol is transmitted. As an example, consider two random variables $W_1\in\{0,1,\dots,N-1\}$ and $W_2\in \{0,1,\dots\}$, and let $U=W_1+W_2N$. Then, the $W_1$ component is always perfectly recovered at the decoder using $(T - 1) \mbox{ mod }N$, as in modulo encoding. On the other hand, the $W_2$ component is estimated as $\lfloor (T-1)/N \rfloor$, which is as close to $W_2$ as can be given $Z_k$.

As a final remark, we note that the Shannon strategies that consider only the current state, i.e., those presented in Section~\ref{sect_shannon}, can also be represented in the timing channel. For example, if the binary Shannon strategies are chosen i.i.d.~with $\mbox{Pr}[U=(0,1)]=p$, then a geometric distributed timing input $V$ with parameter $p$ yields the same channel input distribution and thus the same rate. Similarly, if binary Shannon strategies are chosen by a first order Markov process, an i.i.d.~timing input strategy $U$ that yields the same input distribution can be constructed. Hence, encoding schemes for the timing channel include the Shannon strategy schemes of Section~\ref{sect_shannon}. However, for codebooks generated with higher order Markov processes, it is necessary to have timing auxiliary sequences $U^n$ with memory, and a function $v_k(U_k,Z^k)$ that utilizes the history of the states.

\section{Capacity with Infinite-Sized Battery and No Battery} \label{sect_inf0}

For the purposes of comparison, in this section, we present two extreme cases, the case of no energy storage, and the case of infinite-sized energy storage.

\subsection{Capacity with Zero Energy Storage} \label{sub_inf0_zero}

We first consider an encoder without energy storage capability. That is, we allow a non-zero channel input $X_i=1$ only if energy is harvested within that channel use, i.e., $E_i=1$. We note that this is slightly different than the \emph{transmit first} model described in Section~\ref{sect_model}, where the channel input is sent before energy harvesting in each channel use. In contrast, here we consider a \emph{harvest first} model. For this model, $E_i$ can be considered as an i.i.d.~channel state known at the encoder \cite{ozel2011awgn}, for which the capacity is given in (\ref{eqn_timing_csit}). Using the Shannon strategies $U_1=(0,0)$ and $U_2=(0,1)$, with $\mbox{Pr}[U_2]=p$, the capacity in this case becomes
\begin{align}
\label{eqn_inf0_zero_capacity}
C_{ZS} &= \underset{p}{\max}~ H_2(pq)-pH_2(q)
\end{align}
where $H_2(p)$ is the binary entropy function.

\subsection{Capacity with Infinite Energy Storage} \label{sub_inf0_inf}

Next, we consider the case with an infinite-sized battery at the encoder. Reference \cite{ozel2012achieving} studies the Gaussian counterpart of this channel, showing that the save-and-transmit scheme is optimal. A similar argument applies for the binary case, implying that a rate of $H(X)$ can be achieved, where $X$ is constrained as $\mathbb{E}[X] \leq q$. Hence, the capacity of the channel with an infinite-sized storage is
\begin{equation}
\label{eqn_inf0_infinite_capacity}
C_{IS}=
	\begin{cases}
		H_2(q), & \quad q \leq \tfrac{1}{2} \\
		1, & \quad q > \tfrac{1}{2}
	\end{cases}
\end{equation}

\section{Extension to the Ternary Channel} \label{sect_ternary}

The equivalence of the energy harvesting channel and the timing channel extends beyond binary channels. As an example, in this section, we present results for the ternary energy harvesting channel (TEHC). The TEHC has three input and output symbols, $X,Y \in \{-1,0,1\}$, and both $X=-1$ and $X=1$ require one unit of energy to be transmitted. This extension can further be generalized to $M$-ary channels, with each symbol consuming either 0 or 1 unit of energy.

\subsection {Achievable Rates with Shannon Strategies} \label{sect_ternary_shannon}

In this section, we consider achievable rates with Shannon strategies in the actual channel use index of TEHC. As in the BEHC case, we only have two states, $S_i \in \{0,1\}$. A strategy $U$ is in the form $U=(X,X^\prime)$, where $U(0)=X$ and $U(1)=X^\prime$. Note that $X=1$ or $X=-1$ is possible only when $S=1$, and thus we only have three feasible strategies, namely $(0,0)$, $(0,-1)$ and $(0,1)$.

We first consider codebooks generated by choosing $U_i$ i.i.d.~for each codeword and channel use. Let the probability of choosing $U_i=(0,-1)$ and $U_i=(0,1)$ be $p_2$ and $p_3$, respectively, for all $i$ and all codewords. First, note that this construction yields an ergodic battery state process, with the transition probabilities
\begin{align}
	\mbox{Pr}[S_{i+1}=1|S_i=0]=q, \qquad \mbox{Pr}[S_{i+1}=0|S_i=1]=(p_2+p_3)(1-q)
\end{align}
yielding the stationary probability
\begin{align}
\label{eqn_shannon_naive_prob2}
	\mbox{Pr}[S=1]=\frac{q}{p_2+p_3+q-(p_2+p_3)q}
\end{align}
Note that the stationary probability is a function of $p_2+p_3$, rather than $p_2$ and $p_3$ individually. Denoting $U=(0,0)$ as $0$, $U=(0,-1)$ as $-1$ and $U=(0,1)$ as $1$, the channel in the case of \naive Shannon strategies is expressed as
\begin{align}
\label{eqn_shannon_naive_channel2}
	p(y|u)=\mbox{Pr}[S=1] \delta(y-u)+\mbox{Pr}[S=0]\delta(y)
\end{align}
The best achievable rate with this scheme is given by
\begin{align}
\label{eqn_shannon_naive_rate2}
	R_{NIID}=\underset{p_2,p_3\in[0,1]}{\max} ~
		H(Y)-(p_2+p_3)H_2 \left( \frac{q}{p_2+p_3+q-(p_2+p_3)q} \right)
\end{align}
where $H_2(p)$ is the binary entropy function. We observe that whenever $p_2 + p_3$ is kept constant, the channel in (\ref{eqn_shannon_naive_channel2}) and the term $(p_2+p_3)H_2 \left( \frac{q}{p_2+p_3+q-(p_2+p_3)q} \right)$ in (\ref{eqn_shannon_naive_rate2}) remain unchanged. On the other hand, $H(Y)$ is a concave function of the distribution of $Y$. Hence, by Jensen's inequality, when we fix $p_2+p_3=2p$, selecting $p_2=p_3=p$ yields the highest rate in (\ref{eqn_shannon_naive_rate2}). Therefore, the optimum selection is $p_2=p_3=p$, and we obtain the following simpler rate expression:
\begin{align}
\label{eqn_shannon_naive_rate3}
	R_{NIID}=\underset{p\in[0,1]}{\max} ~
		H(Y)-2pH_2 \left( \frac{q}{2p+q-2pq} \right)
\end{align}

Similar to the BEHC case, the decoder can exploit the memory by using the $n$-letter joint probability $p(u^n,y^n)$ for the channel and obtain {\em optimal i.i.d.~Shannon strategy} (OIID), which achieves the following rate:
\begin{align}
\label{eqn_shannon_oiid_rate2}
R_{OIID}=\max_{p \in [0,1]} \lim_{n \rightarrow \infty} \frac{1}{n}I(U^n;Y^n)
\end{align}
where again $p_2=p_3=p$, whose optimality follows from similar arguments as before. Calculating the limit of the $n$-letter mutual information rate $\frac{1}{n}I(U^n;Y^n)$ is possible by using the algorithm in \cite{arnold2006simulation}. Moreover, we can further improve such achievable rates by constructing codebooks with Markovian Shannon strategies. We evaluate and compare these achievable rates in Section~\ref{sect_numerical}.

\subsection{Timing Equivalence and Related Bounds} \label{sect_ternary_timing}

In order to find a timing equivalent for the TEHC, we represent the channel output $Y^n\in\{-1,0,1\}$ with two sequences, $T^m \in \{1,2,\dots\}^m$ and $L^m \in \{-1,1\}^m$. Here, $T_k$ is the duration between the $(k-1)$st and the $k$th non-zero outputs in $Y^n$, and $L_k$ is the sign of the $k$th non-zero output. As in the binary case, $(T^m,L^m)$ and $Y^n$ are different and complete representations of the same channel output, and therefore are equivalent.

The timing equivalent of the TEHC consists of two parallel channels, namely a timing channel and a sign channel, expressed as
\begin{align}
\label{eqn_ternary_channels}
T_k=V_k+Z_k, \qquad L_k=Q_k
\end{align}
where $Q_k$ is the sign of the $k$th non-zero input. Extending Lemma~\ref{lem_timing_equivalent} to include the sign channel, we observe that the sum capacity of the two independent channels in (\ref{eqn_ternary_channels}) is equal to the capacity of the TEHC. The capacity of the noiseless sign channel is $\log_2|L|=1$ bit per channel use. One use of the sign channel also requires $\mathbb{E}[T]$ uses of the TEHC on average. Considering this, the capacity of the TEHC is given in the following theorem.

\begin{theorem}
The capacity of the ternary energy harvesting channel is
\begin{align}
\label{eqn_ternary_capacity}
	C_{TEHC} = \underset{p(u),v(u,s)}{\max}~ \frac{I(U;T) + 1}{\mathbb{E}[T]}
\end{align}
\end{theorem}

This result is parallel to those in reference \cite{anantharam1996bits} on queues with information-bearing packets. In the timing equivalent of the TEHC, each non-zero channel input can be interpreted as a packet bearing one bit of information. Hence, as in \cite{anantharam1996bits}, coding for the two channels in (\ref{eqn_ternary_channels}) is performed independently, yielding the capacity in (\ref{eqn_ternary_capacity}).

The upper and lower bounds for the BEHC immediately extend to the TEHC, since the capacity for the sign channel is simple. The two upper bounds on $C_{TEHC}$ become
\begin{align}
\label{eqn_ternary_genie}
	C_{UB}^{genie} &=\underset{p \geq 0}{\max}~ \frac{H_2(p)/p+1}{\frac{1}{p}+\frac{1-q}{q}} = \underset{p \geq 0}{\max}~ \frac{qH_2(p)+pq}{q+p(1-q)} \\
\label{eqn_ternary_leakage}
	C_{UB}^{leakage} &= \underset{p_T(t) \in \mathcal{P}}{\max}~ \frac{H(T)-\sum_{t=1}^\infty \frac{H_2((1-q)^t)}{1-(1-q)^t}p(t)+1}{\mathbb{E}[T]}
\end{align}	
where $\mathcal{P}$ is given in (\ref{eqn_ub_leakage_p}), and the two achievable rates become
\begin{align}
\label{eqn_ternary_modulo}
	R_{A}^{mod} &=\underset{N}{\max}~ \underset{p(u),~U\in\{0,1,\dots,N-1\}}{\max}~ \frac{H(U)+1}{\mathbb{E}[V+Z]} \\
\label{eqn_ternary_extended}
	R_{A}^{ext} &=\underset{N}{\max}~ \underset{p(u),~U\in\{0,1,\dots\}}{\max}~ \frac{I(U;Y)+1}{\mathbb{E}[V+Z]}
\end{align}
with $v(U,Z)$ is given in (\ref{eqn_achievable_modulo_v}) for the modulo encoding scheme, and in (\ref{eqn_achievable_extended_v}) for the extended modulo encoding scheme.

\subsection{Capacities with Zero and Infinite Storage} \label{sect_ternary_zero_inf_bat}

We first consider the capacity with zero energy storage. That is, we allow a non-zero channel input $X_i=1$ or $X_i=-1$ only when energy is harvested in that channel use, i.e., $E_i=1$. Using the Shannon strategies $U_1=(0,0)$, $U_2=(0,-1)$ and $U_3=(0,1)$, with $\mbox{Pr}[U_2]=p_2$ and $\mbox{Pr}[U_3]=p_3$, the capacity becomes
\begin{align}
C_{ZS} &= \underset{p_2,p_3}{\max}~ H(Y)-(p_2+p_3)H_2(q)
\end{align}
where $Y$ has the ternary distribution $(p_2 q,1-(p_2+p_3)q,p_3q)$ and $H_2(p)$ is the binary entropy function. Since $H(Y)$ is a concave function of the distribution of $Y$, when $p_2 + p_3$ is fixed, by Jensen's inequality $p=p_2 = p_3$ is the optimal selection. Therefore, we get 
\begin{align}
C_{ZS} &= \underset{p}{\max}~ H(Y)-2pH_2(q) \label{eqn_zero_ter_zero_capacity}
\end{align}
where $Y$ has the distribution $(pq,1-2pq,pq)$.

Next, we consider the capacity with an infinite-sized battery. Similar to the binary case, a rate of $H(X)$ can be achieved, where $X$ is a ternary variable that is constrained as $\mathbb{E}[X^2] \leq q$. Hence, the capacity of the channel with infinite-sized storage is
\begin{equation}
\label{eqn_inf_ter_infinite_capacity}
C_{IS}=
	\begin{cases}
		H(q/2,1-q,q/2), & \quad q \leq \tfrac{2}{3} \\
		\log_2(3), & \quad q > \tfrac{2}{3}
	\end{cases}
\end{equation}
where $H(q/2,1-q,q/2)$ denotes the entropy of the ternary distribution $(q/2,1-q,q/2)$.

\section{Numerical Results} \label{sect_numerical}

In this section, we compare the timing channel upper bounds and achievable rates in Sections~\ref{sect_upperbounds} and \ref{sect_achievable}, Shannon strategy based achievable rates in Section~\ref{sect_shannon} and capacity results for extreme cases in Section~\ref{sect_inf0} for the BEHC, followed by the results in Section~\ref{sect_ternary} for the TEHC. The upper bounds and achievable rates for the BEHC evaluated at $q\in\{0,0.1,\dots,1\}$ are given in Table~\ref{table_values}.

\begin{figure}[t]
\centerline{\includegraphics[width=0.6\linewidth]{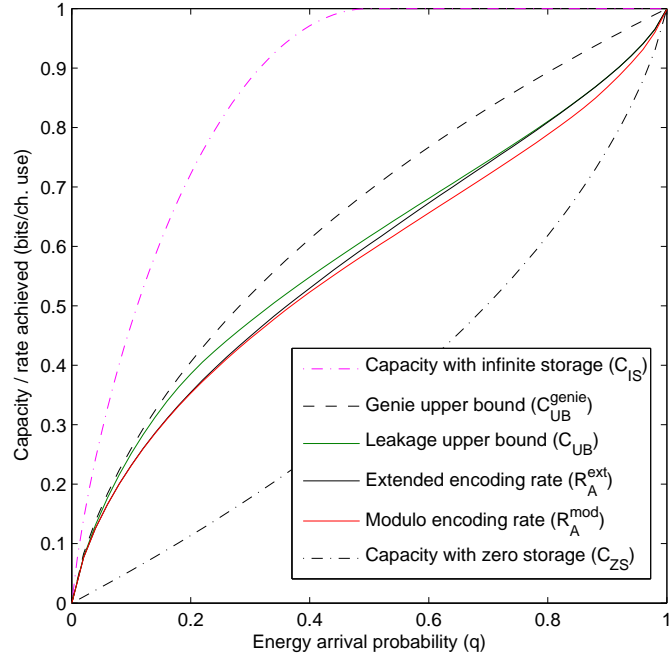}}
\caption{Upper bounds and achievable rates for the BEHC.}
\label{fig_numerical_behc_some}
\end{figure}

Fig.~\ref{fig_numerical_behc_some} shows the genie upper bound $C_{UB}^{genie}$ in (\ref{eqn_ub_genie_final}), the leakage upper bound $C_{UB}^{leakage}$ in (\ref{eqn_ub_leakage_upperbound}), the modulo encoding achievable rate $R_{A}^{mod}$ in (\ref{eqn_achievable_modulo_r}), and the extended encoding achievable rate $R_{A}^{ext}$ in (\ref{eqn_achievable_extended_r}) in comparison with the zero storage capacity $C_{ZS}$ in (\ref{eqn_inf0_zero_capacity}) and the infinite-sized storage capacity $C_{IS}$ in (\ref{eqn_inf0_infinite_capacity}). All of these quantities are zero at $q=0$, because in this case, no energy is harvested, and thus no communication is possible. Moreover, they are all equal to $1$ at $q=1$, because in this case, the battery is always full, and the channel is equivalent to a binary noiseless discrete memoryless channel without any energy constraints. 

From Fig.~\ref{fig_numerical_behc_some}, we first observe that the leakage upper bound, $C_{UB}^{leakage}$, and the achievable rate with the extended encoding scheme, $R_A^{ext}$, provide a gap smaller than 0.03 bits per channel use for the capacity, for all harvesting rates $q$. For small $q$, both upper bounds and both achievable rates get very close, as expected from the asymptotic optimality of $R_{A}^{mod}$ as $q \rightarrow 0$. On the other hand, for large $q$, we observe that the genie upper bound $C_{UB}^{genie}$ is looser compared to the leakage upper bound $C_{UB}^{leakage}$. This implies that the correlation between the harvesting process and the channel outputs is high in this regime. Finally, we note that although the gap between the infinite storage capacity $C_{IS}$ and the zero storage capacity $C_{ZS}$ is large, a unit-sized energy storage device recovers a significant amount of this difference. This demonstrates that even the smallest sized energy storage device can be very beneficial in energy harvesting communication systems.

\begin{table}
\centering
  \begin{tabular}{ | c || c | c | c | c | c | c | c | }
    \hline
    Arrival prob. ($q$)  & $C_{UB}^{genie}$ & $C_{UB}^{leakage}$ & $R_{A}^{ext}$ & $R_{A}^{mod}$ & $R_{M2}$ & $R_{M1}$ & $R_{OIID}$ \\ \hline \hline
0   & 0      & 0      & 0      & 0      & 0      & 0      & 0      \\ \hline
0.1 & 0.2600 & 0.2516 & 0.2317 & 0.2313 & 0.2199 & 0.2188 & 0.2178 \\ \hline
0.2 & 0.4056 & 0.3854 & 0.3546 & 0.3529 & 0.3415 & 0.3384 & 0.3351 \\ \hline
0.3 & 0.5184 & 0.4740 & 0.4487 & 0.4451 & 0.4364 & 0.4320 & 0.4301 \\ \hline
0.4 & 0.6125 & 0.5485 & 0.5297 & 0.5230 & 0.5178 & 0.5130 & 0.5115 \\ \hline
0.5 & 0.6942 & 0.6164 & 0.6033 & 0.5914 & 0.5890 & 0.5880 & 0.5861 \\ \hline
0.6 & 0.7669 & 0.6807 & 0.6729 & 0.6562 & 0.6617 & 0.6591 & 0.6555 \\ \hline
0.7 & 0.8326 & 0.7442 & 0.7403 & 0.7205 & 0.7301 & 0.7301 & 0.7270 \\ \hline
0.8 & 0.8927 & 0.8101 & 0.8088 & 0.7881 & 0.8005 & 0.7997 & 0.7987 \\ \hline
0.9 & 0.9483 & 0.8846 & 0.8845 & 0.8678 & 0.8808 & 0.8807 & 0.8797 \\ \hline
1   & 1 & 1 & 1 & 1 & 1 & 1 & 1 \\ \hline
  \end{tabular}
  \caption{Upper bounds and achievable rates for the BEHC.}
  \label{table_values}
\end{table}

\begin{figure}[t]
\centerline{\includegraphics[width=0.6\linewidth]{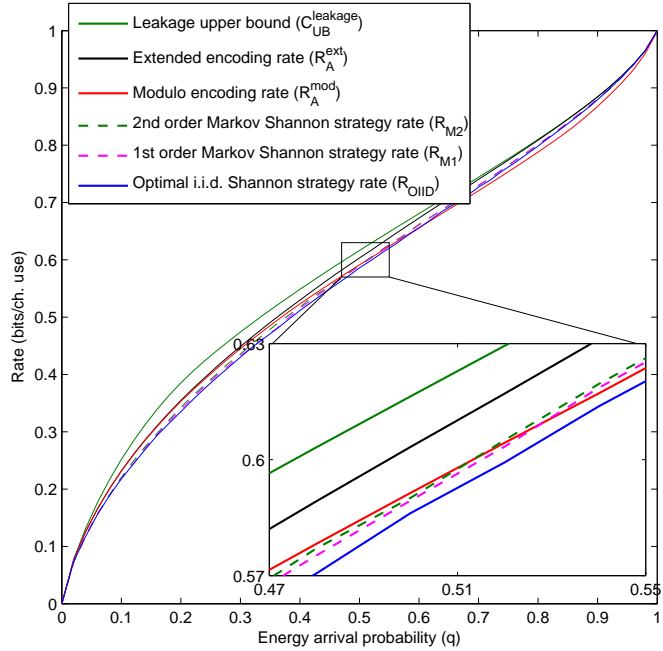}}
\caption{Achievable rates with timing encoding compared with instantaneous Shannon strategies for the BEHC.}
\label{fig_numerical_behc_achievable}
\end{figure}

We next compare the modulo and extended achievable rates, $R_A^{mod}$ and $R_A^{ext}$, with the Shannon strategy based achievable rates described in Section~\ref{sect_shannon}. We remind that the schemes in Section~\ref{sect_shannon}, which are also studied in \cite{mao2013capacity}, only observe the instantaneous battery state in each channel use. Thus, we have simple Shannon strategies, but we allow a Markovian dependence over time in the codewords. Fig.~\ref{fig_numerical_behc_achievable} shows $R_A^{mod}$ and $R_A^{ext}$ along with the optimal i.i.d.~Shannon strategy rate $R_{OIID}$ in (\ref{eqn_shannon_oiid_rate}) and the optimal 1st and 2nd order Markov Shannon strategy rates $R_{M1}$ and $R_{M2}$. We observe that although $R_A^{mod}$ outperforms $R_{OIID}$ for all $q$, the 1st and 2nd order Markov Shannon strategies outperform $R_A^{mod}$ for large $q$, as seen in the inset in Fig.~\ref{fig_numerical_behc_achievable}. However, the extended encoding rate $R_A^{ext}$ outperforms both $R_{M1}$ and $R_{M2}$, for all harvesting rates $q$. These can also be observed partially (for harvesting rates $q\in\{0,0.1,\dots,1\}$) from Table~\ref{table_values}. We note that the increase in the achievable rate with the Markov order of the input seems to be small. However, due to the exponential increase in the computational complexity with the Markov order, it was not tractable to simulate and compare inputs of higher Markov orders, i.e., 3rd and higher Markov orders.

\begin{figure}[t]
\centerline{\includegraphics[width=0.6\linewidth]{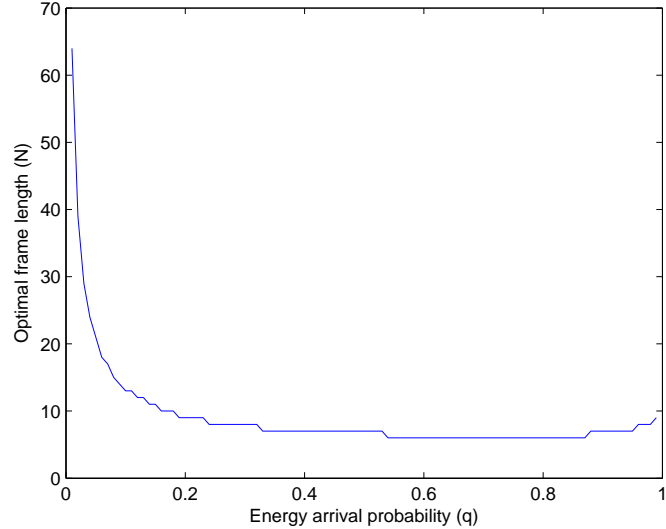}}
\caption{Optimal choice of frame length $N$ for the modulo encoding scheme.}
\label{fig_numerical_behc_Noptimal}
\end{figure}

A parameter of interest is the optimal frame length $N$ for the modulo encoding scheme in Section~\ref{sub_achievable_modulo}, which we present in Fig.~\ref{fig_numerical_behc_Noptimal}. The larger $N$ is, the larger the support of $U$ is, and more information can be packed into a single use of the timing channel. However, as $N$ increases, so does $\mathbb{E}[T]$, and thus each symbol takes more time, and more harvested energy is potentially wasted. Thus, for small harvest rates, e.g., $q \leq 0.7$, optimal $N$ decreases with increasing $q$ so that less harvested energy is wasted. On the other hand, for $q>0.7$, the node is receiving excessive energy, and thus the optimal $N$ increases to pack more information in each timing channel use.

\begin{figure}[t]
\centerline{\includegraphics[width=0.6\linewidth]{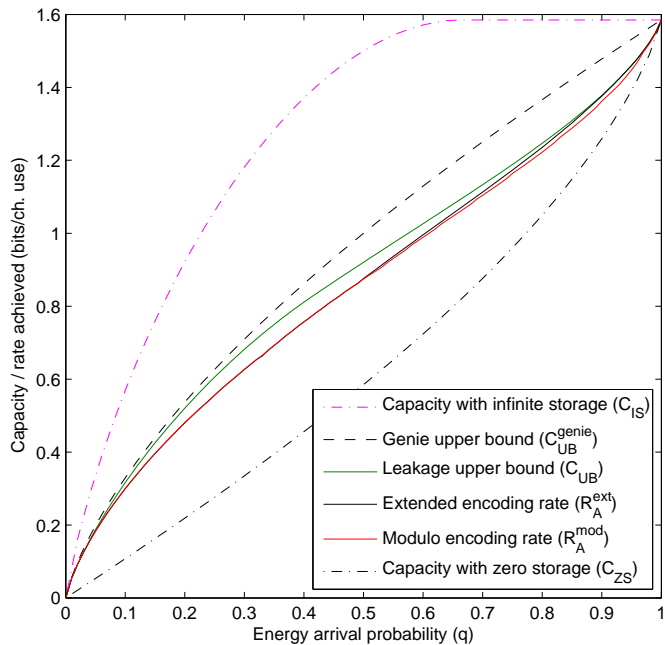}}
\caption{Upper bounds and achievable rates for the TEHC.}
\label{fig_numerical_tehc_some}
\end{figure}

\begin{figure}[t]
\centerline{\includegraphics[width=0.6\linewidth]{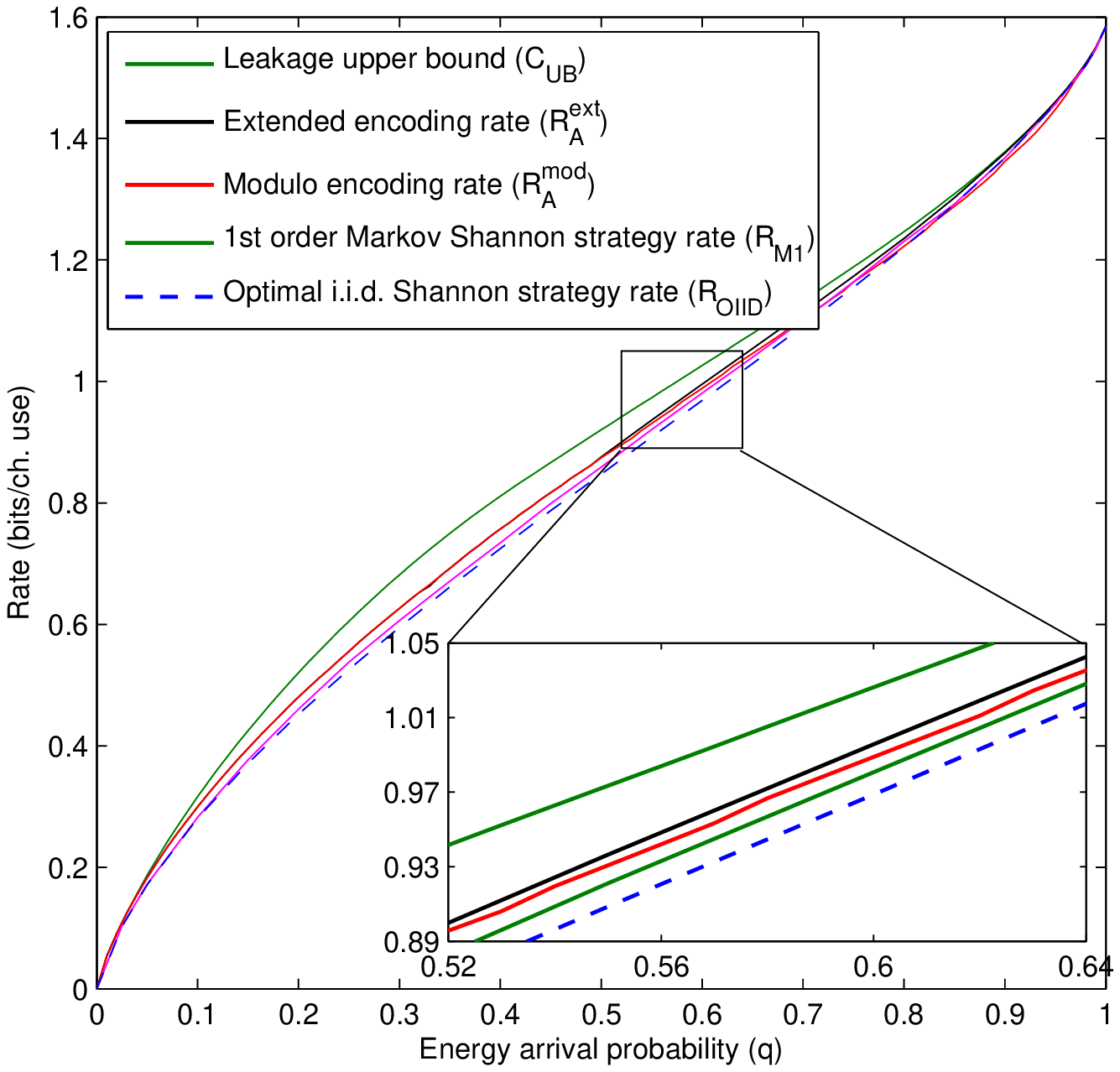}}
\caption{Achievable rates with timing encoding compared with instantaneous Shannon strategies for the TEHC.}
\label{fig_numerical_tehc_achievable}
\end{figure}

Finally, we present the upper bounds and the achievable rates for the ternary channel, given in (\ref{eqn_ternary_genie})-(\ref{eqn_ternary_extended}), together with the zero and infinite-sized battery capacities $C_{ZS}$ and $C_{IS}$ given in (\ref{eqn_zero_ter_zero_capacity})-(\ref{eqn_inf_ter_infinite_capacity}), in Fig.~\ref{fig_numerical_tehc_some}. We also compare the achievable rates in Section~\ref{sect_achievable} with the optimal i.i.d.~and the 1st order Markov Shannon strategies for the ternary channel in Fig.~\ref{fig_numerical_tehc_achievable}. Note that in the ternary channel, the $q=1$ case corresponds to a ternary noiseless discrete memoryless channel, and thus has a capacity of $log_2(3)=1.58$ bits per channel use. We observe that similar to the binary case, the leakage upper bound $C_{UB}^{leakage}$ and the extended encoding rate $R_A^{ext}$ approximate the capacity within 0.05 bits per channel use, and the extended encoding rate outperforms the i.i.d.~and the 1st order Markov Shannon strategies, for all harvesting rates $q$.

\section{Conclusion} \label{sect_conclusion}

Finding the capacity of the binary energy harvesting channel is challenging due to the memory and the input dependence of the battery state. In this paper, we have addressed a simpler case of the binary energy harvesting channel with unit-sized energy storage and without channel noise. For this case, we have shown that the binary channel can also be represented as a timing channel, where the states do not have memory and are not input dependent. Using this equivalence, we have derived two upper bounds: the genie upper bound by providing battery state to the decoder, and the leakage upper bound by quantifying the information leaked to the decoder about energy harvests. We have also proposed two encoding schemes based on a modulo encoding strategy, showing that they are asymptotically optimal for small energy harvesting rates. We have extended these results to the ternary energy harvesting channel. Comparing the upper and lower bounds, we have found the capacities of the binary and ternary energy harvesting channels within 0.03 bits per channel use and 0.05 bits per channel use, respectively. We have also observed that the timing channel based achievable rates outperform i.i.d.~and the 1st and 2nd order Markov Shannon strategies that only consider instantaneous battery states.

\bibliographystyle{unsrt}
\bibliography{bib_eh}

\end{document}